\documentclass[journal]{IEEEtran}

 \usepackage{amsmath,amsthm}
  \usepackage{algorithmic}
  \usepackage{algorithm}
  \usepackage{stmaryrd}
  \usepackage{pxfonts}
  \usepackage{graphicx}
  \usepackage{array}
  \usepackage[nocompress]{cite}

\begin{document}

\title{Resource Allocation in Peer-to-Peer Networks: A Control Theoretical Perspective}

\author{Nitin~Singha,~Ruchir~Gupta,~\IEEEmembership{Member~IEEE,}
        ~Yatindra~Nath~Singh,~\IEEEmembership{Senior~Member,~IEEE}% <-this % stops a space
\thanks{Nitin Singha and Yatindra Nath Singh are with the Department of Electrical Engineering,
	Indian Institute of Technology, Kanpur, India.\protect\\
E-mail:{{nitinss}@iitk.ac.in}}% <-this % stops a space
\thanks{Ruchir Gupta is with Department of Computer Science,
IIITDM Jabalpur }}

\maketitle

\begin{abstract}
P2P system rely on voluntary allocation of resources by its members due to absence of any central controlling authority. This resource allocation can be viewed as classical control problem where feedback is the amount of resource received, which controls the output i.e. the amount of resources shared back to the network by the node. The motivation behind the use of control system in resource allocation is to exploit already existing tools in control theory to improve the overall allocation process and thereby solving the problem of freeriding and whitewashing in the network. At the outset, we have derived the transfer function to model the P2P system. Subsequently, through the simulation results we have shown that transfer function was able to provide optimal value of resource sharing for the peers during the normal as well as high degree of overloading in the network. Thereafter we verified the accuracy of the transfer function derived by comparing its output with the simulated P2P network. To demonstrate how control system reduces free riding it has been shown through simulations how the control systems penalizes the nodes indulging in different levels of freeriding. Our proposed control system shows considerable gain over existing state of art algorithm. This improvement is achieved through PI action of controller. Since low reputation peers usually subvert reputation system by whitewashing. We propose and substantiate a technique modifying transfer function such that systems' sluggishness becomes adaptive in such a way that it encourage genuine new comers to enter network and discourages member peers to whitewash.
\end{abstract}

%% Note that keywords are not normally used for peerreview papers.
%\begin{IEEEkeywords}
%Freeriding, Reputation, Control Theory, Transfer Function, PI Controller, Whitewashing.
%\end{IEEEkeywords}

\section{Introduction}

\IEEEPARstart{I}n recent years, the popularity of the {P}{eer}-to-peer (P2P) system is on the rise, which is  evident from the traffic measurement records of ISP's \cite{eMule}\cite{Kazza}\cite{BitTorrent}. P2P systems follow decentralized network model, where individual users, also referred as peers perform the job of  both server as well as client. Peers share certain portion of their resources  viz. bandwidth, disk storage or processing power with other members. A P2P application combines the resources available across the network and thereafter peers can assist each other in terms of various services like file-sharing \cite{FileSharing}, content lookout \cite{content_search_1}\cite{content_search_2}\cite{content_search_3}\cite{content_search_4}, data storage in distributed manner \cite{DataStorage}, digital content delivery \cite{ContentDelivery}, grid computing \cite{GridComputing} and content transfer \cite{anonymous_transfer}. The term node and peer is used interchangeably in this paper.

The cooperative model of resource exchange in P2P network poses fundamental and challenging research issues. The potential of the P2P system, especially file sharing \cite{FileSharing} and grid computing \cite{GridComputing}, is dependent mainly on the amount of resources contributed by peers. Peers usually are reluctant in sharing resources because there are inherent costs attached with sharing viz. payment for the upload bandwidth. Therefore, natural tendency of peers is to free ride i.e. consume resources without making any contribution back to the network. In worst case, P2P network will fail if nobody shares any resource. Several studies done in past have shown the aforesaid free riding behavior of the member peers in a P2P systems.  Study performed on the Gnutella file sharing system in 2000 \cite{Adar} showed that as many as $70\%$ of its users were free rider and their percentage further increased to 85\% \cite{Hughes} in year $2005$. Nearly half of the total file search responses in network came from 1\% of total nodes. This leads to congestion in these serving nodes thereby leading to the "tragedy of the commons" \cite{tragedyofcommon}. Hence "free riding" and "tragedy of commons" are the two important issues in implementing the cooperative model of information exchange and require immediate attention for increasing the efficiency of the P2P network.

Many solutions such as  micro payment based schemes \cite{karma} \cite{bitcoin}, game theoretic approaches  \cite{Feldman}\cite{MaKiGame}\cite{Huiye} and trust or reputation approaches \cite{TrustEstimation}\cite{Satsiou}\cite{Eigen_Trust}\cite{Peer_Trust}\cite{Power_Trust}\cite{Gossip_Trust} have been proposed in the past which provide incentive to the cooperating peers to solve the above mentioned problems, like. The monetary gain based schemes require   virtual banks or credible authorities for implementation. The mathematical modeling using the game theoretic approaches is  not accurate \cite{Satsiou} because of reliance on some unrealistic assumptions like every node is aware of the link capacity of every other node in the network. Trust or reputation based approaches being simpler and easy to implement are extensively used to encourage the cooperation among users to share their resources in the network. 

In reputation based systems, the incentive is in terms of the service differentiation provided to the requesting node on the basis of its past co-operative behavior. Co-operation level is quantified in terms of peers' reputation. Many resource distribution protocols such as resource bidding mechanism with incentive \cite{MaKiGame}, reputation based resource allocation \cite{Satsiou} and probabilistic resource allocation \cite{probabilistic_allocation} have been proposed in the past, which give preference in service  to the nodes with greater co-operation level. 

The resource allocation by a node based upon the requesters' reputation in the P2P system can be viewed as classical feedback control problem. The output of the control system is the download capacity that overall network offers to the peer. The output of control system follows a set point or reference. In our case, the reference corresponds to an utilization value, which is fixed and equal to $1$. $Utilization = 1$ signifies the optimal case for the peer where it can utilize whole of its download capacity with minimum upload contribution.This proposal is in accordance with tendency of rational peers  \cite{Meo} to adopt a strategy which provides them maximum benefit from the network at minimum cost. Therefore resource allocation at every peer can be implemented using the  control system which always leads to some resource contributes back to network, thereby curbing the peers' free riding tendency. The $Utilization$ is explained in detail in section \ref{SetPoint}. The difference between observed $Utilization$ and the reference is called error. It changes the upload bandwidth offered by the node which subsequently modifies the nodes' reputation, thereby changing the bandwidth offered ot it by the network. The change in the output is always directed such that it minimizes the error. This results in nodes sharing minimum upload capacity to get the required download capacity from the network. 

The proposed control system can also be used to solve the problem of whitewashing \cite{Feldman}\cite{gupta_whitewashing}, where users leave the current system and rejoin it with new identities to avoid low reputation penalties due to their uncooperative behavior in the past and exploit the incentives provided by network to the newcomers on joining the network. Providing no incentive to newcomers will discourage peers to whitewash, however it will also discourage genuine newcomers to join the network. Therefore incentive provided to newcomers is made adaptive, which decreases with increase in whitewashing level. As whitewashing level increases, the control system behaves more sluggish i.e. new entrants require more time to reach optimal state of resource sharing.  

The main purpose of studying the reputation systems  using the control theoretical approach is to use the already existing tools available in the literature in control systems for solving the problem of the free riding and whitewashing.  The main contribution of this paper are listed below 
\begin{enumerate}
	\item We demonstrate, how the whole P2P system can be approximated as linear time-varying model for the purpose of performance control. We also derive and describe in detail the various components of the overall control system viz.~controller,~actuator,~plant and monitor.
	\item  A resource allocation mechanism  is implemented using the control system.  At every node, control system automates the whole process of maintaining a minimum upload bandwidth for getting the requisite download bandwidth from the network.
	\item The transfer function of the proposed control system has been derived. The results obtained from it clearly show that classical control theory can address the problem of the free riding. 
	\item The overall network performance under varying degree of loading at the serving nodes has also been studied and simulation results clearly show that the proposed control system is able to handle these changes.
	\item The equivalent transfer function and the original network with the control system were simulated independently and the results were compared to validate the accuracy of the derived transfer function. The simulation results clearly verify that the transfer function closely models the network in steady state. Acceptable variations were observed during transient state which can be attributed to the linearizion \cite{Ogata} and other assumptions taken during derivation of transfer function. For details please refer to section \ref{The Controlled Software System Model}.
	\item It has been shown through simulation results sharing less than what is instructed by the control system are unable to utilize their complete download capacity. Therefore they they are at loss and consequently don't freeride.
	\item The resource allocation performance with control system  and then  with the existing resource allocation algorithm (R.A.) \cite{Satsiou} was compared. In simulation results it has been observed that the network with control system is able to achieve steady state faster and control system appreciably reduces the steady state error (i.e. difference between the actually shared and the optimal upload capacity). This happens due to the PI controller present in control system.
	%\item The reputation of all the nodes in the network achieves the desired convergence behavior. (This is proved by the stability analysis of the control system).
	\item Finally, the proposed control system has been shown to considerably reduce the whitewashing in the network. First, we propose a method to determine whitewashing level. Afterwards depending upon the current whitewashing level, the transfer function updates itself which changes the speed at which a peer achieves the optimal level of resource sharing i.e. $Utilization =1$.
\end{enumerate}
Rest of the paper is organized as follows. The section \ref{Related_Work} summarizes and compares the related work, whereas the system model is described in section \ref{nwk_model}. In section \ref{Reputation_model}, reputation management system along with the resource allocation is discussed in detail. In section \ref{C_sys}, we model the overall P2P network as control system and derive its transfer function. The section \ref{Performance_Evaluation} presents performance evaluation of the proposed control system in achieving optimality and controlling free riding, simulation results to verify accuracy of transfer function and finally comparison with the existing (R.A.) \cite{Satsiou} allocation algorithm. Discussion about whitewashing problem and its solution using control system is provided in section \ref{Whitewashing}. Finally paper concludes with section \ref{conclusion}.

 
%\hfill September 17, 2014

\section{Related Work}
\label{Related_Work}
Resource allocation strategies to prevent free riding have been proposed by many authors in the past. Satsiou and Tassiulas \cite{Satsiou} put forward Reputation based resource allocation (R.A.), which gives preference to the requesting nodes having higher reputation during the allocation. Higher reputation implies greater contribution level, therefore R.A. strategy motivates users to co-operate. In R.A. strategy requesting peers are sorted by the corresponding serving peer in the decreasing order of their reputation to demand ratio. The serving peer, first satisfies the request of the peer highest in the list, afterwards if the resources are left then it will be distributed to the next highest peer in the list and so on. A peer increases or decreases its capacity in fixed step sizes to manipulate its reputation based on the amount of the resource it wants from the overall network. The model in \cite{Satsiou} is limited to the single capacity link i.e. sum of the upload and download bandwidth will be constant and nodes will decide the partitioning of bandwidth between upload and download. 
Gupta \emph{et al.} proposed probabilistic resource allocation \cite{probabilistic_allocation} where nodes are selected probabilistically for resource allocation. Probability of the requesting node getting selected for the resource allocation increases with increase in its reputation. However the lower reputation nodes still have some finite probability of getting selected and therefore they don't get completely disconnected from the allocating node.
In both  R.A. \cite{Satsiou} and probabilistic resource allocation \cite{probabilistic_allocation},  the bandwidth is increased or decreased  in fixed step size therefore it is very difficult to achieve the convergence of the upload bandwidth. During the steady state upload bandwidth will keep on oscillating with amplitude of oscillation proportional to the step size. Transient response of the network can be improved if the step size is kept variable. Both the problems mentioned above are addressed and described in detail in section \ref{comparison_RA}.

Among the other related work \cite{Meo} Meo \emph{et al.}, contemplated network as market and proposed that second auction strategy leads to the optimality. However the study is limited to the homogeneous peers with the single capacity. Another serious limitation with this scheme is that the member node can perform only single upload and download at a given time. Ma \emph{et al.} \cite{MaKiGame} put forward a modified version of the progressive water filling algorithm, where requesting nodes are provided resources at the rates proportional to their reputation. Ma \emph{et al.} made an analogy of the resource allocation with filling a bucket, where base of the bucket is directly proportional to the reputation of the node. The author theoretically  proved that such kind of allocation maximizes the marginal utility. However author has not supported his hypothesis with simulation analysis. Yan \emph{et al.} \cite{Yan} presented a theoretical framework for the optimal resource allocation. Peers' ranking and the utility are the basis for the resource allocation such that resource distribution achieves max-min fairness. The theoretical analysis done in  \cite{MaKiGame}\cite{Yan} considers only stationary competing peers. Dynamics of competing peers like variation in the demand of requesters is absent in their study.

Various papers have suggested array of techniques in past to prevent free riding in P2P network. Direct reciprocity schemes like "tit for tat" \cite{tit_for_tat} employed in Bit Torrent file systems \cite{Cohen} encourage co-operative behavior among the member peers. In direct reciprocity schemes service provided by the node $i$ to the node $j$ will solely depend on service $j$ has provided to $i$ in the past, while completely ignoring $j$'s contribution to the other users in the system. However such kind of schemes  are effective when members remain in the system for long time so that there are ample opportunities for reciprocation among them. In the current P2P networks where there is large population size, high churn rate \cite{Stutzbach} and infrequent repeat transactions, the reputation systems are more effective than the direct reciprocity schemes. Feldman \emph{et al.} \cite{Feldman} gave the concept of the generosity of a node, which is the ratio of the service provided to the service received by the node from the network. Based upon the estimated generosity, the node will be imparted service from the network. However no mathematical analysis was provided by Feldman \emph{et al.} for countering the problem of the free riding in the P2P network. In the same context Kung \emph{et al.} \cite{Kung} advocated that nodes' usage of the resources and contribution back to the network should be the basis of the selection of the nodes for the resource distribution. Nodes are expected to contribute above a certain threshold level to be eligible for the resource allocation. Banerjee \emph{et al.} \cite{Banerjee} proposed that serving peer will compute the expected utility function of the requesting nodes. The decision to share by the serving peer will be based upon this expected utility function. Gupta \emph{et al.} \cite{TrustEstimation} put forward a reputation based system that takes into account various uncertainties at the serving node to arrive at more accurate estimate of the reputation. It also carries out the exponential moving average of the present and past reputation values to provide appropriate weightage to the past behavior of the node. However solution proposed in \cite{Feldman}\cite{Kung}\cite{Banerjee}\cite{TrustEstimation} do not discuss about mechanism to distribute resources among the requesting nodes. 

Friedaman \emph{et al.} \cite{Friedman} pointed out that whitewashing takes place in P2P networks due to availability of the free identities or cheap pseudonyms. Many authors \cite{Castro}\cite{Izhak-Ratzin} in the past have suggested that whitewasher in a network can be completely removed by assigning permanent identities to its members. However John R. Doucher \cite{Sybil_Attack} asserted  that it is possible for members in decentralized network like P2P networks to have multiple identities. Therefore P2P network suffers from the problem of free riding. Yang \emph{et al.} \cite{Yang_Whitewashing} studied Maze file sharing system and came to conclusion that incentives encourage whitewashing. Friedaman \emph{et al.} \cite{Friedman} proved that considering every new comer as a potential whitewasher is always a better policy over any other static stranger policy for avoiding whitewashing. Grolimund \emph{et al.} \cite{Grolimund} proposed to assign low initial reputation to the new comers in the network for preventing free riding. However assigning low reputation to newcomers will discourage legitimate newcomers from joining the network. Therefore initial reputation needs to be adjusted periodically on the basis of level of whitewashing so that there are enough incentives which motivate new users to join the network and at the same time degree of whitewashing can be reduced.

A control system is proposed in this paper to avoid free riding for both stationary and dynamically competing peers. It employs PI controller \cite{Ogata} which reduces the steady state error of the P2P system and at same time helps the system to achieve steady state faster. On modifying the initial reputation (which is granted to the newcomer at the time of joining the network) on the basis of the level of whitewashing, the control system can successfully counter the problem of the whitewashing in the network.

\section{P2P Network Model}
\label{nwk_model}
\begin{figure}
	\centering
	\includegraphics[width=0.7\linewidth]{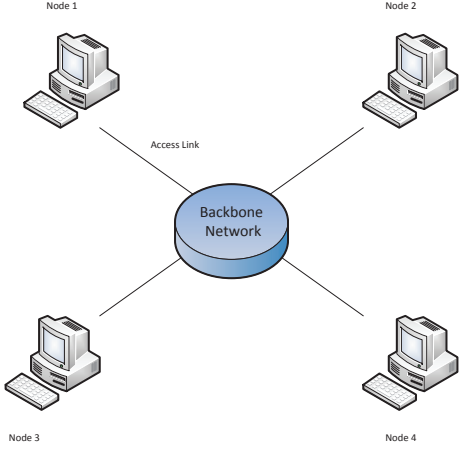}
	\caption{The P2P Network Model}
	\label{p2pnetwork}
\end{figure}

We analyze a unstructured P2P network of N nodes, where nodes consume and share the bandwidth on demand. There is no centralized server and every peer acts as both client and server concurrently. Each peer in the network is connected to backbone network through an access link. In this way peers are connected to each other as shown in Fig. \ref{p2pnetwork}.

In this paper we study an overloaded network as considered in \cite{Satsiou}\cite{RationalMalicious} where at any given node, there are always requests pending to download data and content are present for downloading. %The number of requests  being generated and requests being handled by the node in a given times slot is maximum due to overloading across the network %.
Segmentation of content into chunks and their distribution across the network is out of scope for this paper.

The network is assumed to consist of symmetric links only i.e. for a link between two nodes $i$ and $j$, the feasible data rate and the bandwidth required to support it, is same  in both the directions. Download requirements at the node $i$ are assumed greater than or equal to maximum download capacity of the node $i$. This assumption is taken to model the worst case  of maximum overloading in the network. It is further assumed that the total upload bandwidth  available at the node $i$ is sufficient, such that when it is bartered with other requesting nodes in the network, the download requirement at the node $i$ will always be fulfilled. 

For the purpose of modeling, time is divided into discrete time slot. Each time slot signifies a period. We have considered unstructured P2P network for simulation but the proposed control system is applicable for structured P2P networks also. However we have not performed any simulations  to substantiate this claim. Every time slot is termed as a round. The nodes send the requests at the starting of the round and if they gets selected for the service allocation then their request are fulfilled by the serving nodes within the same round. For the next round the whole process is repeated again.  

\section{Reputation model}
\label{Reputation_model}
Any P2P network is meaningful only when nodes cooperate and fulfill each others' requirements. Network under consideration, contains selfish peers who are rational i.e they are interested in maximizing their own benefit without caring about the other members. Selfish nodes never want to contribute back to the network. Therefore they will always try to become a free rider, unless their is some incentive attached in sharing of their resources. This leads to the overloading at the few nodes which are contributing back to the network also known as \emph{tragedy of commons}. In the worst case, the network will collapse  when  every node becomes a free rider. For avoiding the problems of \emph{free riding} and \emph{tragedy of commons}, a reputation management system can be used in P2P network. Every node's reputation is evaluated at the end of the each round based upon its performance in that round. 	
\subsection{Reputation Calculation}
The trust is measure of the cooperative behaviour of the node. The trust model used in this paper is similar to the one proposed by Satsiou and Tassiulas \cite{Satsiou}. The trust calculated by the requesting node $j$ for the serving node $i$  during the $n^{th}$ period is defined as the ratio of the bandwidth ($r_{ji}^n$) received by the node $j$ to what it has actually demanded ($B_{ji}^n$) from the node $i$  during the $n^{th}$ period i.e.
\begin{equation}
	t_{ji}^n=\frac{r_{ji}^n}{B_{ji}^n}.
\end{equation} 
For ease of reading, the important notations are listed in Table~\ref{rep_tab}.
\begin{table}[]
	\caption{Notations with their Description}
	\centering
	\begin{tabular}{ |c|c|}
		\hline
		\textbf{Notation}  &\textbf{Description} \\
		
		\hline
		$R_{i}^n$ &Reputation of node $i$  at the end of  $n^{th}$ time slot\\
		\hline
		$t_{ji}^n$ &Trust of $i$ as measured by  $j$ for  $n^{th}$ time slot\\
		\hline
		$r_{ji}^n$ &Bandwidth received by $j$ from $i$ for $n^{th}$ time slot\\
		\hline
		$B_{ji}^n$ &Bandwidth demanded  by $j$ from  $i$ for $n^{th}$ time slot\\
		\hline
		$B_i^{max}$ &The maximum download capacity of the node $i$  \\
		\hline
		$l_i^n$ &No. of requests received by the node $i$ in $n^{th}$ slot \\ 
		\hline
		$g_i^n$ &No. of requests generated by the node $i$ in $n^{th}$ slot \\ 
		\hline
		$B_{ij}^{fes}$ &The minimum amount of bandwidth required to  \\
		& support feasible data rate practically possible on the  \\   
		& link between $i$ and $j$\\
		\hline
		$B^{fes}_{min}$ &The minimum value of the $B_{ji}^{fes}$, among all the\\ 
		&   possible values of $i$ and $j$ in the network \\ 
		\hline
		$k_{{ij}_{ovd}}^{n}$ &Proportionality constant for handling change in \\
		& bandwidth received by $i$ due to overloading at $j$\\ 
		\hline
		$A^{n}_i$   & Set of serving nodes for the node $i$ during $n^{th}$ period\\
		\hline
		$S^{n}_i$   & Set of requesting nodes, requesting resources from \\
		
		&   node $i$ during $n^{th}$ period\\
		\hline
		${R_{min}}$ & Reputation threshold below which node does not \\
		& receive any service from network\\
		\hline
		$W^n$         & Level of Whitewashing during $n^{th}$ period\\
		\hline
		$N^n$      & Total nodes in network during $n^{th}$ period\\
		\hline
		${R_{in}^{max}}$ & Maximum initial reputation that can be assigned \\ 
		& to newcomers\\
		\hline 
		${R_{in}^n}$  &  Initial reputation assigned to newcomers during \\
		& $n^{th}$ period\\  
		\hline
	\end{tabular}
	\label{rep_tab}
\end{table}       
Reputation of the node $i$ for the $(n+1)^{st}$ time slot is calculated using the exponential moving average \cite{EMA} of reputation up to the $n^{th}$ period with the average of trust values for the current time period as give below
\begin{equation}
	\label{reput_g}
	R_i^{n+1}= (\alpha)R_i^n+(1-\alpha)\frac{\sum\limits_{j \in S_i^{n+1}}t_{ji}^{n+1}}{\sum\limits_{j \in S_i^{n+1}}1}.
\end{equation}
In the above equation ${R_i}^{n+1}$ is the reputation of the node $i$ at the end of the $(n+1)^{st}$ period and $S_{j}^{n+1}$ is the set of nodes which have requested the node $i$ for resources during $(n+1)^{st}$ period. The cardinality of the set $S_{j}^{n+1}$ i.e ($\sum\limits_{j \in S_i^n}1$) is equal to $l_i^{n+1}$. Here $l_i^{n+1}$ is the number of the requests received by the node $i$ during the $(n+1)^{st}$ period. Using reputation aggregation algorithm \cite{Gossip_Trust}\cite{DifferentialGossip},  reputation tables are updated periodically by every node so as to maintain identical reputation tables across the network. $\alpha$ determines how the exponential average forgets the contribution of the past values and maintains its value in between $[0,1]$. Higher $\alpha$ discounts the older observations faster and $\alpha=1$ completely ignores the  older values. We take value of $\alpha$ as $\frac{1}{2}$ in the equation (\ref{reput_g}) so that equal\footnote{In this way nodes are discouraged to abruptly change their behavior during one time period so as to gain advantage in the reputation for the future transactions. The nodes need to be co-operative over a period of time for getting better quality of services than the existing services.} priority is given to the present as well as past reputation values.Therefore the reputation of the node $i$ in this paper is calculated as
\begin{equation}
	\label{reput}
	R_i^{n+1}= \frac{R_i^n+\frac{\sum\limits_{j \in S_i^{n+1}}t_{ji}}{l_i^{n+1}}}{2}.
\end{equation}
The reputation of a node is the measure of its cooperative behavior. The service or bandwidth allocated to a node in a network increases with the reputation of the node and vice versa. Therefore for acquiring more download capacity from the network, the node has to share more bandwidth towards the network. 
\subsection{Reputation Based Resource Distribution }
\label{allocation}
The resources are distributed among the requesting nodes using weighted water filling algorithm \cite{MaKiGame}. In our proposed model, the reputation of the node is used as its weight during the resource assignment. The source node  distributes the resources among the requesting nodes at the same time but with rates equal to the their reputation.  If reputation of a requesting node is below a certain threshold say $R_{min}$ then it will not be allocated any resource. This is done to discourage malicious behavior of nodes. During the resource allocation, a requesting node $i$ will be taken out of allocation process if either resources at the source gets exhausted or assigned resources to $i$ become equal to what $i$ had demanded for the current round. Let $\hat{1}, \hat{2}....\hat{M}$ be the M requesting nodes sorted in non-decreasing order of their $\frac{b_k}{R_k}$. $R_k$  and $b_k$ are the reputation and bandwidth requested by the node $k$. The bandwidth allocation is done in the order as per the sorted list. Let $\hat{r}_{k}$ denote the bandwidth allocated to node $k$. It is given as
\begin{equation}
	\label{allc}
	\hat{r}_{k}=min\left\{b_{k},\frac{R_{k}(W_s-\sum\limits_{i=1}^{k-1}\hat{r}_i)}{\sum\limits_{j=k}^{M}R_j}\right\}; \qquad k=\hat{1},....,\hat{M}.
\end{equation} 
Here $W_s$ is the total upload bandwidth of the serving node. Suppose $\overrightarrow{b}=[1,1.5,2,5]$ is the bandwidth demand of the requesting nodes in $Mb/s$ from the source with upload capacity of $7Mb/s$. Let the reputations of the requesting nodes be  $\overrightarrow{R}=[0.625,0.25,0.625,1]$, then the bandwidth allocated in $Mb/s$ using weighted water filling algorithm is given by $\overrightarrow{r}=[1,0.8,2,3.2]$. 

The bandwidth allocated to any requesting node $i$ from $j$ during the $(n+1)^{st}$ time period is proportional to its reputation $R^{n}_i$ in the  previous period\footnote{The reputation of the requesting node during current round will be evaluated at the end of time period and will be used as weight in the next round.}, bandwidth demanded $B_{ij}^{n+1}$ from $j$ and a proportionality constant $k_{{ij}_{ovd}}^{n+1}$ which caters for change in received bandwidth due to overloading at $j$.
Therefore bandwidth received by $i$ from $j$ during $(n+1)^{st}$ period is given
\begin{equation}
	r_{ij}^{n+1}=k_{{ij}_{ovd}}^{n+1}{\times}R^{n}_i{\times}B_{ij}^{n+1}.
\end{equation}
The value of $k_{{ij}_{ovd}}^{n+1}$ lies in between $[0,\frac{1}{R^{n}_i}]$. $k_{{ij}_{ovd}}^{n+1}=0$ denotes the worst case of overloading at serving node $j$, where the receiving node $i$ will not receive any bandwidth. When $k_{{ij}_{ovd}}^{n+1}=\frac{1}{R^{n}_i}$ then the bandwidth requirement of node $i$ will be completely satisfied by the node $j$. The value of proportionality constant $k_{{ij}_{ovd}}$ varies in each round depending upon the degree of the overloading at the serving node. At the same time for another requester say $x$, requesting the same serving node, the value of $k_{{xj}_{ovd}}$ will vary depending upon its reputations.

The total bandwidth received by the node $i$ during $(n+1)^{st}$ period is given by
\begin{equation}
	\label{capacity}
	C_i(n+1)=R_i^{n}\times{\sum\limits_{j\in{A^{n+1}_i}}{k_{{ij}_{ovd}}^{n+1}}B_{ij}^{n+1}},
\end{equation}
where $A^{n+1}_i$ is the set of the nodes from which node $i$ has demanded resources during the $(n+1)^{st}$ round. We assume reputation aggregation so that ${R_i}^n$ estimated is same for all the nodes.
\section{P2P network modeling as Control System}
\label{C_sys}
The objective of the control system is to enable each peer to maintain the minimum upload so as to maintain the just enough  reputation, to get the desired download capacity. The control theoretic modeling can improve the performance of the reputation management system as a peer can adapt and derive maximum utility from the network. The design of a control system  involves determining the equivalent components i.e. controller, actuator, plant and monitoring unit. A mathematical model has been developed for this purpose. Finally the overall performance of the devised control system model need to be evaluated to determine the amount of the improvement achieved. The additional overhead incurred in this process also needs to be carefully analyzed. The following subsections discusses the set point determination and  various other components which constitute this control system. 

\subsection{Determination of the Set Point}
\label{SetPoint}
The proportion of the  requesting peer $i's$ demand which is fulfilled by the network in the $n^{th}$ round in comparison to its download bandwidth is given by 
\begin{equation}
	\label{satisfaction}
	{
		U_i(n)=\frac{C_i(n)}{B_i^{max}}.
	}
\end{equation}
Here $C_i(n)$  is the total data rate agreed by all the serving peers to be given to the requesting node $i$  and  $B_i^{max}$ is the maximum download capacity of the requesting node $i$ from the nodes in the network .The serving peers will never give wrong information about the $C_i(n)$ because their reputation is calculated on the basis of actual data rate delivered and not the data rate they are willing to provide i.e. $C_i(n)$. We assume that all the peers are rational, therefore there are no malicious peer. When a peer is assumed  rational then it will  not act maliciously \cite{RationalMalicious}, as rational peers gain nothing by harming other peers. Therefore a node will not receive incorrect $C_i(n)$.

When the requesting node $i$ increases its upload capacity then its reputation increases consequently making $i$ eligible for more download capacity $C_i(n)$ from the network. Hence $C_i(n)$ of the requesting node increases. Three cases may arise in the equation (\ref{satisfaction}). When $C_i(n)<B_i^{max}$ then satisfaction derived by the peer $i$ from the network increases with increase in $C_i(n)$. The satisfaction attains its maximum values at $C_i(n)=B_i^{max}$ i.e. $U_i(n)=1$. After $U_i(n)=1$, higher value of $C_i(n)$ will actually not be useful to the node because it exceeds the maximum data rate the requesting node can handle.  Node is at loss when $U_i(n)>1$ because it has to pay more in terms of the greater upload capacity for the download rate limited by $B_i^{max}$. Therefore to derive the maximum benefit from the network, a node should maintain $U_i(n)=1$, while keeping the upload rate to minimum, so that it can full-fill all of its download requirement at the minimum cost.

Hence the set point or reference point for our control system is $U_{ref}=1$. The feedback monitor in the loop will estimate function $U_i(n)$; this value will be compared with the reference point ($U_{ref}=1$) and the error is fed to the controller. It is assumed in the system model proposed in section \ref{p2pnetwork}, that every node $i$ has adequate upload capacity, such that when it is bartered with other requesting nodes in the network, then download requirement of  node $i$ will always be satisfied. In other words upload capacity available  at node  is adequate enough to make the $U_i(n)$ at the node $i$ equal to the $U_{ref}$ of the control system.

The $U_i(n)$ discussed above is different from the concept of utility  \cite{Shenker}. The utility corresponds to the degree of the satisfaction received by the node from the network whereas $U_i(n)$ is dependent upon the total resources that are available to the node from the network. Some time the nodes' capacity may not be sufficient to utilize all the resources. Therefore these extra resources which are considered in $U_i(n)$ are useless for the node and don't contribute in overall satisfaction and consequently in the utility derived by the node. 
\subsection{The Controller}
The controller \cite{Ogata} stabilizes the system output to a particular value called set point. It compares the difference between the desired and the actual $U$ in the proposed control system. Based upon this comparison, the controller drives the actuator to regulate the output. Generally PID controller is preferred as a controller in control system. However when the P2P network reaches the steady state, nodes generally share in proportion to their requirement so as to preserve their reputation. Therefore total shared capacity in network does not change abruptly and consequently so does $U$. The differential action is used to counter sudden changes in output thus it is not required in our system. Therefore the PI controller with transfer function $G(s)=K_{p}(1+\frac{k_i}{s})$ \cite{Ogata} is used  in this paper to model the control system. $K_{p}$ and $K_{i}$ are  the proportional and integral gain respectively  of the PI controller.
\subsection{The Actuator}
\label{act}
The role of the actuator \cite{Ogata} in the control loop is to update physical entity based upon the controller output. In our case, the actuator modifies the upload capacity of node in response to the controller output $y$. If $y$ is the input to the actuator then $y\Delta$ is the the upload capacity shared by the node in the next round. The change in the upload capacity by the actuator modifies the reputation of the node, thereby adjusting the download data rate, the node receives from the network. $\Delta=\frac{{B_i}^{max}}{10}$ is taken in accordance with the existing allocation algorithms \cite{Satsiou}\cite{probabilistic_allocation} for ease of comparison with them.

%If the value of the $\Delta$ is large, then the node reaches the desired reference utility very  quickly but the steady state error will be high and vice versa. The value of  $\Delta$ can be determined from the simulation results. We have to look for the point where a good trade off can be reached between convergence rate and performance of the system. We can assume $\Delta = \frac{1}{10}U_{ref}$.    

%Therefore node $i$'s actuator keeps on dynamically updating its $\Delta$ value depending upon its current state depicted by the $U_i$. The current value of $\Delta$ is equal to some proportion of the maximum possible change of the value of the $U_i$. The proportionality is inversely proportional to how close the $U_i$ is to the desired reference point\footnote{As we reach closer to the reference point we want to reduce the steady state error therefore value of $\Delta$ should be small.}. Therefore the value of $\Delta$ is calculated as
%

%\begin{equation}
%\Delta=|U_{ref}-U_i|_{max}\times|U_{ref}-U_{i}(n)|.
%\end{equation} 
%Here $U_i(n)$ corresponds to the value of the $U_i$ for the current $n^{th}$ round.  $|U_{ref}-U_i|_{max}=1-0=1$. Note that the values of $U_i$ on the side $U_i>1$ will be very less. This is because of the overloading in the network so $C_i(n)$ never reaches 2$B_i^{max}$. 
%
%Therefore for control system proposed in the paper, $\Delta$ becomes
%\begin{equation}
%\Delta=|U_{ref}-U_i(n)|.
%\end{equation}   
%Another way around is that $\Delta = 5 \% of (U_{ref})$. This can be a valid choice as band of 5\% corresponds to the settling time in the control system. 
\subsection{The Monitor}
The function of the monitor at the node $i$ is to sense the current data rate $C_i(n)$ offered to the node $i$, for evaluating the $U_i(n)$. The evaluated $U_i(n)$ is compared with the set point to calculate the error and drive the controller which in turn decides how much resources are shared by the node $i$ to the other nodes. Let $(y\Delta)_{ji}^{n+1}$ corresponds to amount of resources allocated by the node $i$ to the requesting node $j$ during  $(n+1)^{st}$ round. The trust calculated by the requesting node $j$ for the serving node $i$ in the $(n+1)^{st}$ round is given by
\begin{equation}
	\nonumber
	t_{ji}^{n+1}=\frac{(y\Delta)_{ji}^{n+1}}{B_{ji}^{n+1}}.
\end{equation}
Using equation (\ref{reput}) the reputation of the node after the $(n+1)^{st}$ period is given by 
\begin{equation}
	\label{Rep}
	%R_{i}^{n+1}=\frac{mR_{i}^{n}+{\frac{y{\Delta}{l_i^{n+1}}}{{l_i^{n+1}}{B_{min}}}}}{m+l_i(n+ %1)},     \nonumber \\
	%{\sum\limits_{j\in{A^{n+2}_i}}B_{ij}(n+1)}
	R_{i}^{n+1}=\frac{R_i^n+\frac{{\sum\limits_{j\in{S^{n+1}_i}}{\frac{(y\Delta)_{ji}^{n+1}
					}{B_{ji}^{n+1}}}}} {l_i^{n+1}}}{2}.
\end{equation}
The symbols have been defined along with the equation (\ref{reput}). The total bandwidth received by the node $i$  during $(n+2)^{th}$ time period  is calculated using equation (\ref{capacity}) as 
\begin{equation}
	\nonumber
	C_i(n+2)=R_{i}^{n+1}\times{\sum\limits_{k\in{A^{n+2}_i}}k_{{ik}_{ovd}}^{n+2}{\times}B_{ik}^{n+2}}.
\end{equation}
Putting the value of $R_i^{n+1}$ from equation (\ref{Rep}) into the above equation we get
\begin{equation}
	\label{C_i}
	C_i(n+2)=\frac{R_i^n+{\frac{\sum\limits_{j\in{S^{n+1}_i}}\frac{(y\Delta)_{ji}^{n+1}}{B_{ji}^{n+1}}}{l_i^{n+1}}}}{2}\times{\sum\limits_{k\in{A^{n+2}_i}}k_{{ik}_{ovd}}^{n+2}{\times}B_{ik}^{n+2}}.
\end{equation}
The  serving peers  allocate resources in proportion to the demands of the requesting peers.The requesting peers may receive much less bandwidth then what they had requested because of the overloading at the serving peer. In order to optimize for themselves, the requesting peers will demands more than their actual capacity. Normally the requesting peer should be requesting whatever is its download capacity. To overcome the problem of nodes demanding more than their actual capacity, the serving peers can estimate the feasible capacity of the links to the requesting peers and use it instead of what is demanded if the demand is more than the feasible capacity. Feasible capacity is the minimum amount of the bandwidth required to support the feasible service rate across the link. Feasible service rate is the maximum achievable throughput via underlying path in the network with packet loss probability
$p$. It can be estimated for the TCP Reno algorithm \cite{feasible_rate} as
\begin{equation}
	\nonumber
	\label{feasible_rate}
	R(p){\approx}\frac{M}{RTT*\sqrt{\frac{2bp}{3}}+T_{0}Min(1,3\sqrt{\frac{3bp}{8}})p(1+32p^2)}.
\end{equation}
Here $R(p)$ is the feasible service rate which is the function of the packet loss probability $p$, $M$ is the maximum transmit window that the receiver indicates to the sender and RTT the round trip time between the two nodes. $T_0$ is the retransmission time out in seconds and $b$ is the number of packets acknowledged by each acknowledgment message.  

The serving peer will never give more than feasible bandwidth as it knows  that remaining bandwidth will be wasted.  Hence the equation (\ref{C_i}) becomes 
\begin{gather}
	\nonumber
	C_i(n+2)=\frac{R_i^n+{\frac{\sum\limits_{j\in{S^{n+1}_i}}\frac{(y\Delta)_{ji}^{n+1}}{B_{ji}^{fes}}}{l_i^{n+1}}}}{2}\times{\sum\limits_{k\in{A^{n+2}_i}}k_{{ik}_{ovd}}^{n+2}{\times}B_{i}^{max}}\\
	=\frac{R_i^n+{\frac{\sum\limits_{j\in{S^{n+1}_i}}\frac{(y\Delta)_{ji}^{n+1}}{B_{ji}^{fes}}}{l_i^{n+1}}}}{2}\times{B_i^{max}}{\sum\limits_{k\in{A^{n+2}_i}}k_{{ik}_{ovd}}^{n+2}}.
\end{gather}
Here $B_i^{max}$ is the maximum download capacity of the node $i$ and $B_{ij}^{fes}$ is the minimum amount of the bandwidth required to support the feasible data rate, as derived from the equation (\ref{feasible_rate}), on the link between node $i$ and $j$. We assume symmetric links in the system model in the  section \ref{nwk_model}. Therefore $B_{ji}^{fes}=B_{ij}^{fes}$ i.e feasible data rate while transferring data from node $i$ to $j$ is same as that from $j$ to $i$.

%${\sum\limits_{j\in{A^{n+2}_i}}1}$, represents the cardinality of the set $A^{n+2}_i$ and is equal to $g_i^{n+2}$, the number of the nodes which have been requested for the resources by the node $i$ during $(n+1)^{st}$ time slot. Therefore equation (\ref{C2}) reduces to
%\begin{equation}
%C_i(n+2)=\frac{R_i^n+{\frac{\sum\limits_{j\in{S^{n+1}_i}}\frac{(y\Delta)_{ji}^{n+1}}{B_{ji}^{fes}}}{l_i^{n+1}}}}{2}\times (B_i^{max})({g_i^{n+2}}).
%\end{equation}
Using equation (\ref{satisfaction}), the measured variable $U_i$ corresponding to the node $i$ for the $(n+2)^{th}$ time period is given by
\begin{equation}
	\label{utility_1}
	U_i(n+2)=\frac{C_i(n+2)}{B_i^{max}}=\frac{{R_{i}^{n}}+{\frac{\sum\limits_{j\in{S^{n+1}_i}}\frac{(y\Delta)_{ji}^{n+1}}{B_{ji}^{fes}}}{l_i^{n+1}}}}{2}\times{\sum\limits_{k\in{A^{n+2}_i}}k_{{ik}_{ovd}}^{n+2}}.
\end{equation}
%As peers are rational so to maximize their chances of getting content from the network they try to generate maximum amount of data request. Therefore $g_i^{n+2}=g_{max}$, the maximum amount data requests generated by the node $i$ for the given time slot.
%Hence $U_i(n+2)$ becomes 
%\begin{equation}
%\label{utility_2}
%U_i(n+2)=\frac{{R_{i}^{n}}+{\frac{\sum\limits_{j\in{S^{n+1}_i}}\frac{(y\Delta)_{ji}^{n+1}}{B_{ji}^{fes}}}{l_i^{n+1}}}}{2}\times{g_{max}}.
%\end{equation}
\subsection{The Controlled Software System Model}
\label{The Controlled Software System Model}
\begin{figure}
	\centering
	\includegraphics[scale=0.45]{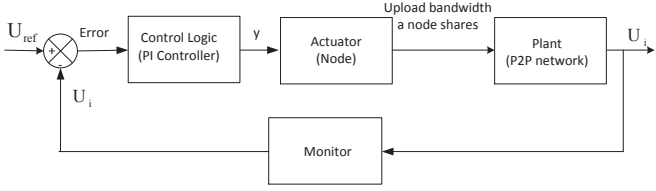}
	\caption{The Block Diagram of Control System at the Peer}
	%\caption{The control loop of a peer}
	\label{control block}
\end{figure}
\begin{figure}
	\centering
	\includegraphics[scale=0.5]{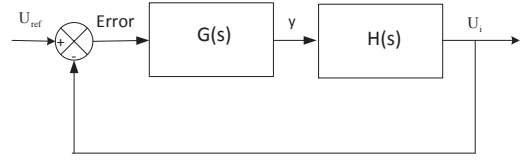}
	\caption{The control loop }
	\label{Transfer_Function}
\end{figure}

  
Fig. \ref{control block} shows the block diagram of the control system which measures the data rate received by the node and maintains it at the desired level by regulating the upload bandwidth of the node. Let $H(s)$ denotes the controlled software system (which includes actuator and P2P network) as shown in Fig. \ref{Transfer_Function}. For simplicity we neglect the process dynamics\footnote{Process dynamics play significant part during transient phase of the system. We assume network has reached steady state and its behaviour is virtually constant with time} and consequently $H(s)$ is equal to static gain $h$. On linearizing \footnote{If any non linear system involves small signals and operates around the equilibrium point then it can be approximated as linear system within a limited range of the operation.}, the gain $h$ \cite{Ogata} around the reference point $(U_{ref}=1)$, we obtain gain as the  derivative of the output ($U_i$) with respect to the input $(y)$ as shown in the Fig. \ref{control block}. Using equation (\ref{utility_1}), the static gain of the requesting node $i$ is given by 
\begin{equation}
	\label{S_gain}
	h=\frac{dU_i(n+2)}{dy}=\frac{d\left[\left(\frac{{R_{i}^{n}}+{\frac{\sum\limits_{j\in{S^{n+1}_i}}\frac{(y\Delta)_{ji}^{n+1}}{B_{ji}^{fes}}}{l_i^{n+1}}}}{2}\right)\times{\sum\limits_{k\in{A^{n+2}_i}}k_{{ik}_{ovd}}^{n+2}}\right]}{dy}.
\end{equation}
The maximum gain\footnote{The system is designed for the maximum gain. If system is stable for the maximum gain then it will also be stable for the lesser.} $h_{max}$ occurs when $l_i^{n+1}=1$ i.e. node $i$ receives only one request during given time slot and $B_{ji}^{fes}=B^{fes}_{min}$ . Where $B^{fes}_{min}$ is the minimum value of the $B_{ji}^{fes}$, among all the possible values of $i$ and $j$, i.e. 
\begin{equation}
	\nonumber
	B^{fes}_{min}= \min_{i,j} \;\; B_{ji}^{fes}.
\end{equation}

Another factor $k_{{ik}_{ovd}}^{n+2}$ as discussed in section \ref{allocation} attains its maximum value when reputation of node is minimum. $R_{min}$ is the minimum value of the reputation at which a node is eligible for resource allocation in the P2P network .
Therefore the maximum possible value of ${(k_{{ik}_{ovd}}^{n+2}}$) such that node can receive data from network is $\frac{1}{R_{min}}$.  Hence the equation (\ref{S_gain}) reduces to
\begin{equation}
	\label{max_gain}
	h_{max}=\frac{d\left[\left(\frac{R_{i}^{n}+\frac{(y\Delta)}{B_{min}^{fes}}}{2}\right)\times{\frac{1}{R_{min}}}\times{\sum\limits_{k\in{A^{n+2}_i}}1}\right]}{dy}.
\end{equation}
${\sum\limits_{k\in{A^{n+2}_i}}1}$, represents the cardinality of the set $A^{n+2}_i$ and is equal to $g_i^{n+2}$, the number of the nodes from which resources have been requested by the node $i$ during $(n+2)^{th}$ time slot. As peers are rational so to maximize their chances of getting content from the network they will try to generate maximum amount of data requests hence $g_i^{n+2}=g_{max}$ . Therefore equation (\ref{max_gain}) becomes
\begin{equation}
	\nonumber
	h_{max}=\frac{d\left(\frac{R_{i}^{n}+\frac{(y\Delta)}{B_{min}^{fes}}}{2}\times\frac{1}{R_{min}}\times{g_{max}}\right)}{dy}.
\end{equation}

The maximum static gain $h_{max}$ on differentiation is given as
\begin{equation}
	\label{Sgain}
	h_{max}=\frac{\Delta{g_{max}}}{2{R_{min}}B^{fes}_{min}}.
\end{equation}
Change in upload capacity induced by actuator is observed in the next period. This results in  dead time (delay) of one period $T$.  Therefore the overall transfer function of the controlled system is given by
\begin{equation}
	\nonumber
	\label{control software}
	{H(s)=h_{max}.e^{-sT}}.
\end{equation}
The equation (\ref{Sgain}) does not contain any term related to the node $i$ so it is a generic expression applicable to any of the receiving node. Thus, the transfer function of the controlled software system is given by
\begin{equation}
	H(s)=\left(\frac{\Delta{g_{max}}}{2R_{min}B^{fes}_{min}}\right)e^{-sT}.
\end{equation}

The overall transfer function of the control loop as shown in Fig. \ref{Transfer_Function} is given by 
\begin{equation}
	\label{T(s)}
	T(s)=H(s)G(s)=h_{max}.e^{-sT}K_p(1+\frac{k_i}{s})
\end{equation} 
\subsubsection{Deriving Model Parameters} 
When the frequency of the operation is the natural frequency of oscillation, then the transfer function\footnote{We substitute $s=jw$ in the overall transfer function of the control loop } of the control loop given by equation (\ref{T(s)}) becomes
\begin{equation}
	\label{Ngain}
	h_{max}.e^{-jw_nT}K_p(1+\frac{K_i}{jw_n})=-1.
\end{equation}
The gain margin $G$ specifies the stability of the control system and is set by the designer. Therefore using equation (\ref{Ngain}), the gain and the phase equations for closed loop at natural frequency becomes
\begin{equation}
	\label{S1}
	h_{max}|(e^{-jw_nT})(K_p)(1+\frac{K_i}{jw_n})|=\frac{1}{G},
\end{equation}
and
\begin{equation}
	\label{S2}
	w_nT+tan^{-1}(\frac{K_i}{w_n})=\pi.
\end{equation}
In industrial PI controller it is common practice to tune the controller phase i.e $-tan^{-1}(\frac{K_i}{jw_n})$ to $-\frac{\pi}{6}$ \cite{Pcontrol}\cite{cntrol}, hence taking
\begin{equation}
	\label{S3}
	tan^{-1}(\frac{K_i}{w_n})=\frac{\pi}{6}.
\end{equation}
The parameter $G$ and $T$ are set by the designer and $h_{max}$ is calculated using equation (\ref{Sgain}). %Therefore three equations (\ref{S1}), (\ref{S2}), (\ref{S3}) contain three unknowns and can be easily solved to get the value of the parameters $K_p$, $K_i$ and $w_n$. 
Using equation (\ref{S3}) in (\ref{S2}) we get the value of $w_n$ as
\begin{equation}
	\nonumber
	w_n=\frac{5\pi}{6T}=\frac{2.618}{T}rad/s.
\end{equation} 
The parameter $K_i$ of controller is calculated by substituting above computed value of $w_n$ in equation (\ref{S3}). The simple algebraic manipulations gives
\begin{equation}
	\nonumber
	K_i=\frac{1.512}{T}.
\end{equation}
The above computed values of $w_n$ and $K_i$ are used in equation (\ref{S1}) for obtaining the expression for $K_p$ as
\begin{equation}
	\nonumber
	K_p=\frac{0.866}{h_{max}G}.
\end{equation}
\begin{table}[]
	\caption{Simulation Parameters along with their Values}
	\centering
	\begin{tabular}{ |c|c|c|}
		\hline
		\textbf{Parameter}  &\textbf{Description} &\textbf{Value} \\
		\hline
		&Number of the nodes in   &$100$ \\ 
		$N^n$    &network  during $n^{th}$ period & \\ 
		\hline
		$K_p$   &Proportional Gain     &$3.4641\times10^{-3}$ \\
		\hline
		$K_i$   &Integral Gain     &$1.51149$ \\
		\hline
		$G$   &Gain Margin     &$10$ \\
		\hline
		&Maximum number of  requests   & \\
		$g_{max}$ &generated by the node in     &$2$ \\
		& a given time period     & \\
		\hline
		$\alpha$   &Exponential moving     &$\frac{1}{2}$ \\
		&average constant & \\
		\hline
		& Maximum initial reputation that can  &0.01\\ 
		${R_{in}^{max}}$               & be assigned to newcomers &  \\
		\hline 
		&  Initial reputation assigned to   &\\
		& newcomers during $n^{th}$ period, except  &0.01  \\  
		${R_{in}^n}$       &  during  whitewashing scenario , &\\
		&   where its value is calculated   & \\
		& using  equation (\ref{whitewash}) &\\
		
		\hline
		& Reputation threshold below which &\\
		& node does not receive any service  &\\
		 ${R_{min}}$   & from network, except for  &0.01\\
		&  whitewashing scenario  simulation,  &\\
		&  where its value is equal to  ${R_{in}^n}$ &\\
		\hline
		$\Delta$   &Step Size     &$0.5$Mbps \\
		\hline
		&  The minimum amount of bandwidth    & \\
		& required to support feasible data rate  & \\   
		$B_{min}^{fes}$              & practically possible on the link&2Mbps\\
		&  between nodes $i$ and $j$ &\\
		\hline
	\end{tabular}
	\label{sim_tab}
\end{table}       
\section{Performance Evaluation}
\label{Performance_Evaluation}
The current section examines the performance of proposed control framework through simulations.The various simulation parameters along with their values are listed under table \ref{sim_tab}. The values of parameters $K_p$, $K_i$, and $\alpha$ are evaluated from the equations given in this paper and parameter $G's$ values is taken from \cite{cntrol} and the remaining parameters value are taken from \cite{Satsiou} for the ease of comparison. During simulation time is taken as discrete and each discrete time slot signifies a period. The network gets initialized  at $time ~T=10$. Before $time ~T=10$ the $Utilization$ of every node present in network is $0$. The simulation analysis of the transfer function, actual P2P network and the RA \cite{Satsiou} an existing state of art algorithm has been undertaken in this section.
\subsection{Study of Control System Performance}
\begin{figure}[!t]
	\centering
	\includegraphics[scale=0.45]{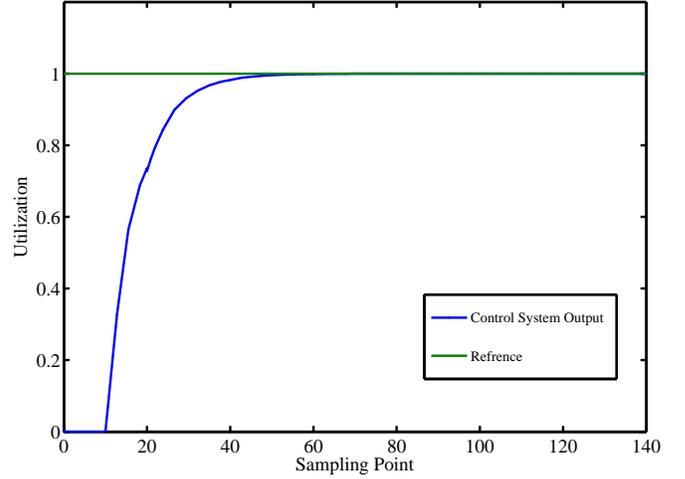}
	\caption{Control System performance in a P2P network}
	\label{Control_Performance}
\end{figure}

Fig. \ref{Control_Performance} demonstrate how the control system is able to assist a node  in the P2P network in attaining a stable value of the utilization  around the reference point i.e $U_{ref}=1$. Initially the utilization of a node will be $0$ as it has not shared anything, afterwards it changes and  stabilizes around $U_{ref}$. To achieve the target utilization (i.e $U_{ref}$) the node adjusts its upload bandwidth, which causes the change in the reputation of the node further leading to the change in the download bandwidth received by the node. The experimental results show that above discussed adjustments pushes the utilization of the node to the target utilization.

%Fig. \ref{Variable_GM} compares utilization variation (with different value of the gain margins i.e. $G=4$ and $G=10$). Fig. \ref{Variable_GM} clearly shows that when G is less, the system is less stable and stability increases with increase in G. However as $G$ increases the system becomes more sluggish. Therefore an optimal value of gain margin is required for system operation.

%\begin{figure}[!t]
%\begin{center}
%\includegraphics[scale=0.45]{Gain_MArgin.eps}
%\caption{Gain Margin variations on control system}
%\label{Variable_GM}
%\end{center}
%\end{figure}

\begin{figure}[!t]
	\begin{center}
		\includegraphics[scale=0.45]{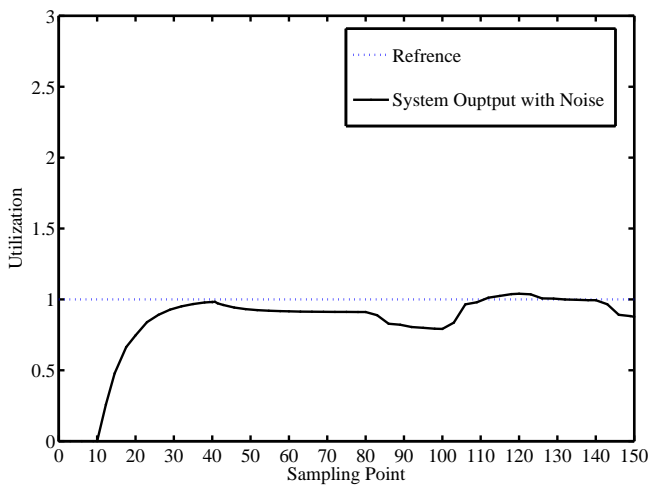}
		\caption{System performance under artificially induced loading in P2P network}
		\label{ServiceRate}
	\end{center}
\end{figure}

When P2P network enters steady almost every user has settled down in terms of demand, therefore requirement of the user doesn't change abruptly. Consequently download capacity received by a node for a given amount of shared upload capacity remains almost constant. This fact is also demonstrated by the simulation results in Fig. \ref{Control_Performance} where $Utilization$ remains around $1$ during steady state operation. For checking the performance of proposed control system during worse case scenario we artificially induce high level of loading in the network during steady state. %change in loading between $[-.25,.25] around the actual demand received during the steady state. The positive value of change in loading implies overloading and negative value signify underloading. 
 Loading at serving node is determined by number of requests coming to a serving node. Overloading implies that amount of bandwidth demanded from the serving node is more than the what it has shared. Due to variation in loading the total download capacity promised by the serving node changes with time. The amount of bandwidth received by a node is directly proportional to the download capacity promised by the serving node. %These variations happen due to the change in the load at the serving node. Sometime  serving nodes may have enough bandwidth to fulfill the demand of the receiving node and in  the other case the level of 
Therefore increase in the level of loading at the server results in the receiving node getting promise of lesser bandwidth and consequently smaller resources for the same level of the reputation.
Simulation results in Fig. \ref{ServiceRate} clearly demonstrate that fluctuations in the level of loading at the serving node is successfully handled by the proposed control system model. %The requesting node varies its upload bandwidth which leads to change in its reputation to maintain its utility around $U_{ref}$.
\subsection{Veracity evaluation of the Transfer Function}
\label{TF}

\begin{figure}[!t]
	\begin{center}
		\includegraphics[scale=0.43]{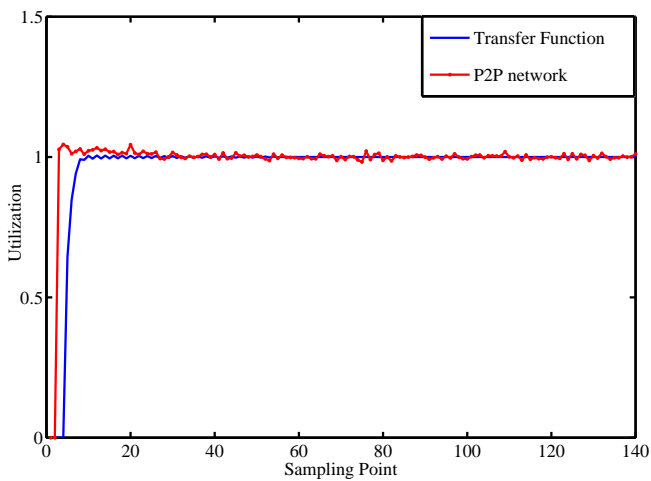}
		\caption{Transfer Function comparison with actual network }
		\label{transfer_function}
	\end{center}
\end{figure}

To verify and show that the derived transfer function in section \ref{The Controlled Software System Model} closely models the proposed P2P network, we simulate the various components of the control system i.e. controller, actuator, plant and monitor. Thereafter results obtained for both the systems discussed above are compared and presented in Fig. \ref{transfer_function}. 

While deriving transfer function in section \ref{The Controlled Software System Model}, we assumed that system has already attained steady state. Therefore the steady state response is same for the both, whereas there is slight variation in the transient response. We had considered worst case during the derivation of transfer function i.e. maximum overloading at the serving node. This results in a sluggish transient response of the transfer function compared to the actual system.

These small deviations in the systems' transient response obtained by the transfer function, from the actual system's response can be tolerated because almost all  P2P systems most of the time operate in steady state. At the same time, without linearizion \cite{Ogata} (i.e assuming system operation in steady state), actual system analysis becomes very complex due to consideration of non-linearities. For considering non-reality, we need detailed information about the network parameters such as link capacities. Consequently it will be very difficult to obtain a generalized control system model for the P2P network.  

\section{Whitewashing}
\label{Whitewashing}
Whitewashing in a network can be reduced by providing low initial reputation \cite{Grolimund} to the newcomers. However this will also discourage newcomers from joining the network.  To make P2P system lucrative for newcomers, we need to increase initial reputation at the cost of higher level of whitewashing. The contradictory requirement for both high level of initial reputation for newcomers and low level for controlling whitewashing can be taken care of by making initial reputation adaptive i.e. it gets modified on the basis of the level of whitewashing in the network. When level of the whitewashing in the network is very high, we can decrease the initial reputation to counter whitewashing. If level of whitewashing is low than initial reputation to newcomers can be increased so as to motivate new comers for joining the network. 

The level of whitewashing in the network is calculated by observing the departing nodes. There are two types of departing nodes in the network. One who have reputation greater than or equal to the initial reputation provided to the newcomers in the network. These nodes will definitely not whitewash because they do not gain any advantage by whitewashing. The remaining departing nodes whose reputation is less than the initial reputation provided to new comers are the prospective whitewashers. We define level of whitewashing $W^{n-1}$ during $(n-1)^{th}$ period as the number of departing nodes in that period which can be possible whitewashers (i.e. have their current reputation less than the initial reputation). 
The initial reputation during the $n^{th}$ period, on the basis of level of whitewashing can be calculated as follows
\begin{equation}
	\label{whitewash}
	{R_{in}^n}=\left(1-\frac{W^{n-1}}{N^{n-1}}\right){R_{in}^{max}}
\end{equation}
where $N^{n-1}$ are the number of nodes during $(n-1)^{th}$ period. The value of $R_{in}^{max}$ is taken as $0.01$ based on earlier work \cite{Satsiou} on whitewashing.

\begin{figure}[!t]
	\begin{center}
		\includegraphics[scale=0.45]{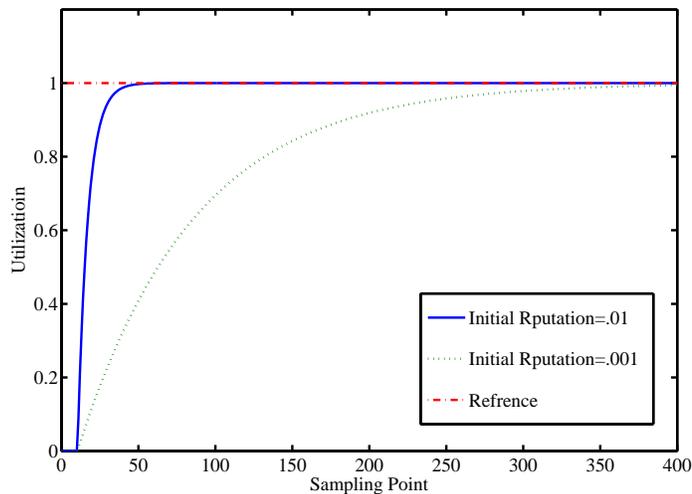}
		\caption{System performance for different level of whitewashing }
		\label{whitewashing}
	\end{center}
\end{figure}

To discourage nodes from leaving the network, the threshold below which node becomes ineligible for service i.e. $R_{min}$ is made equal to the initial reputation of the joining node (${R_{in}^n}$).  The reputation threshold is calculated at the beginning of every period. The change in $R_{min}$ modifies the gain $h_{max}$ of the controller as shown by equation (\ref{Sgain}). %$h_{max}$ can't be manipulated by the node because various parameters required for its calculation can't be changed by the node. 
The node if disregard the formulae and parameter value, then can choose different $h_{max}$ value, though it may not be optimal for it. From the Fig. \ref{whitewashing}, it is evident that as the initial reputation awarded to the node is made smaller, the system becomes more sluggish while discouraging the nodes to leave the network. Sluggishness signifies delay incurred by a node in reaching the optimal level of the resource sharing. More sluggish the system is, greater the time for which the newcomer nodes' $Utilization$ remain less than $1$. If $Utilization <1$, then the node will be at loss, because it will be getting  lesser resources from the network than what it can actually consume. However if white washing level is lower in the network than the value of the initial reputation awarded can be increased so that system becomes faster and genuine new nodes entering the system are not unnecessarily penalized. 

\section{Conclusions}
\label{conclusion}
In this paper, we have used the control theory to solve the problem of freeriding and whitewashing in the P2P networks. We have derived the transfer function of a P2P network which provides a more structured and efficient approach to analyse the P2P networks through use of already existing tools in the control theory. We were able to automate the whole resource allocation process using the control system. We have also demonstrated through simulation that the proposed system is able to help the nodes in sharing optimal level of resources even in the case of high degree of overload in the network. The nodes sharing less than what control system recommends are at loss because they are not able to utilize their total download capacity which discourages the nodes to freeride. Further, we have used PI controller which improves the overall allocation process. The proportional action helps a node in reaching optimal value of resource sharing faster, whereas integral action reduces the steady state error. These improvements are clearly visible when compared with the already existing RA \cite{Satsiou} algorithm. Finally, through the simulations, it has been demonstrated that proposed system can also be used to control whitewashing.

%It will be interesting to study the performance of system in presence of nonlinearities.

% or


%
%\appendices
%% you can choose not to have a title for an appendix
%% if you want by leaving the argument blank
%\section{}
%% use section* for acknowledgment
%\section*{Acknowledgment}

\ifCLASSOPTIONcaptionsoff
  \newpage
\fi



%\bibliographystyle{IEEEtran}
\bibliographystyle{IEEEtran}
\bibliography{ref}


%\end{IEEEbiography}


%\newpage



%\vfill


\end{document}